\documentstyle[12pt]{article}
\begin{document}
\sloppy
\thispagestyle{empty}

\mbox{}\\
NTZ 42/96\\
DESY 96-239\\
hep-ph/9611452\\
November 1996\\

\vspace*{\fill}
\begin{center}
{\LARGE {\bf The evolution of the nonsinglet twist-3 } } \\

\vspace{2mm}
{\LARGE{\bf parton distribution function\footnotetext[1]{
    An earlier version of this paper has been presented at the
  ``3rd Meeting on the Prospects of Nucleon-Nucleon Spin Physics at HERA'',
 JINR Dubna, 28-29.6.1996.
             }
}}\\

\vspace{2em}
{\large B. Geyer, D. M\"uller}
\\
\vspace{2em}
{\it Fakult\"at f\"ur Physik und Geowissenschaft der Universit\"at Leipzig,}
 \\
{\it  Augustusplatz, D-04109 Leipzig,
    Germany}\\
{\ } \\
\vspace{2em}
{\large D. Robaschik}
\\
\vspace{2em}
{\it DESY-IfH,}
 \\
{\it  Platanenallee 6 , D-15738 Zeuthen,
    Germany}\\
\end{center}
\vspace*{\fill}
\begin{abstract}
\noindent
The  twist  three  contributions  to  the $Q^2$-evolution of the
spin-dependent  structure function $g_2(x_{Bj},Q^2)$ are considered in
the  non-local  operator  product approach. Defining appropriate twist
three  distribution function we derive their evolution equation for the
nonsinglet case in leading order approximation. In the limit $x_{Bj}\to 1$
as  well as in the large $N_c$ limit we confirm the result that
the  evolution of the nonsinglet part of $g_2$ is governed by a
Gribov-Lipatov-Altarelli-Parisi equation. \end{abstract}

\vspace*{\fill}
\newpage
\section{Introduction}
Recently,  in  deep  inelastic  scattering  the  first  moments of the
polarized structure function $g_2(x_{Bj},Q^2)$ are measured \cite{g2}.
In  leading  order of the momentum transfer $Q^2$ this structure
function   is   determined  by  twist  two  as  well  as  twist  three
contributions. In comparison with the twist two case the leading order
analysis for the twist three part is more subtle due to the appearance
of  a set of operators mixing under renormalization and constrained by
relations between themselves.

Up to now there exist already several papers which determine the local
anomalous  dimensions  \cite{BKL,JI,KO}  and the evolution kernels for
the   distribution   functions   as   well   as   nonlocal   operators
\cite{BKL,RAT,BB}.  In  the  nonsinglet  sector  the  one-loop  result
\cite{BB}   for  the  evolution  kernel  of  light-ray  operators  was
confirmed  in  \cite{GeyMueRob96}.  It has been checked that the local
anomalous dimensions coincide with the results given in \cite{BKL,KO}.

The  renormalization  properties of twist three operators indicate that
the   evolution   equation  for  the  distribution  function  is  more
complicated  than  the Gribov-Lipatov-Altarelli-Parisi (GLAP) equation
govering  the  evolution  of  twist two parton distribution functions.
However,  for  the  nonsinglet  case  the solution of this equation is
known   in   both   limits   $N_c\to   \infty$   and  $x_{Bj}  \to  1$
\cite{AliBraHil91}.

Here  we  introduce a new twist three distribution function defined in
terms  of three-particle operators and we give first results about the
evolution  in the flavor nonsinglet sector. This distribution function
depends  on  an effective momentum fraction and on the position of the
gluon  field.  The evolution is governed by an extended GLAP equation.
The  moments are derived from twist three operators with definite spin
and  so  that  they  do  not mix with each other. The remaining mixing
problem  can  be treated numerically for the first few moments. In the
limit  $N_c \to \infty$ this equation reduces to an evolution equation
of the GLAP type.

\section{Nonlocal operator product expansion}

The antisymmetric part  (with respect to Lorentz indices) of the hadronic
tensor
\begin{eqnarray}
\label{defHadTen}
W^A_{\mu\nu}&=&
   \epsilon_{\mu\nu\lambda\delta} q^\lambda
   \left[
    {S^\delta\over qP} g_1(x_{Bj},Q^2) +
    {qP\,S^\delta -qS\,P^\delta \over (qP)^2} g_2(x_{Bj},Q^2)
   \right]
\end{eqnarray}
being relevant for polarized deep inelastic scattering is given by the
structure functions $g_1(x_{Bj},Q^2)$ and $g_2(x_{Bj},Q^2)$ depending
on the Bjorken variable $x_{Bj}$ and the momentum transfer $q,\; q^2 =
- Q^2 $. The polarization vector of the proton is defined as
$S^\delta=\bar{u}(P,S)\gamma^\delta\gamma^5u(P,S)$. Our conventions are
$\gamma^5=i \gamma^0\gamma^1\gamma^2\gamma^3$ and $\epsilon_{0123}=1$.
The hadronic tensor is determined  by the absorptive part
$W_{\mu,\nu}$ of the virtual forward Compton amplitude $T^{\mu\nu}$,
\begin{eqnarray}
\label{dispersionrel}
T^{\mu\nu} = {i\over \pi} \int d^4 x e^{ixq}
    \langle P,S|T j^\mu(x/2)j^\nu(-x/2) |P,S\rangle .
\end{eqnarray}
In  the  Bjorken  region  the  leading  terms  in $Q^2$ of the Compton
amplitude  correspond  to  the leading light-cone singularities in the
coordinate  space  so  that  the  light-cone  expansion  for  operator
products  (OPE) can be applied. This provides the factorization of the
structure function in a perturbatively determined coefficient function
and a nonperturbative parton distribution function.

The leading order analysis starts with a perturbative investigation of
the   time  ordered  product  of  two  electromagnetic  currents.
(In the following  we  neglect flavor and color indices):
\begin{eqnarray}
\label{p1.4}
 \{Tj^\mu(x) j^\nu(y)\}^{\rm as} =
    \epsilon^{\mu\nu\rho\sigma} i\partial_\rho^{x} D^c(x,y)\;
    {\bar\psi}(x)\gamma_\sigma\gamma_5 U(x,y)
    \psi(y) + (x,\mu \leftrightarrow y\nu) + \cdots,
\end{eqnarray}
where the path ordered phase factor
$      U(x,y) = P \exp\left\{-ig\int_y^x dw^\mu  A_{\mu}(w)\right\}$
ensures gauge invariance. For simplicity we set $ y=0 $.

Technically,  we apply the light-cone expansion \cite{AZ,FORT}
proved by Anikin and Zavialov in renormalized quantum field theory.
Heuristically, this nonlocal OPE is obtained by approximating
the  vector  $x$  by  the  light-like  vector  $\tilde  x$,
defined as $x=\tilde{x} + a(x,\eta) \eta $, where $\eta$ denotes a fixed
auxiliary vector (e.g. normalized by  $\eta^2  = 1$) and $a$ the
corresponding coefficient.
In leading order we substitute $x \rightarrow \tilde{x} $
whereas  the  $x^2$-singularities  remain  unchanged, i.e.,  $x^2
\rightarrow  x^2$. In general, the coefficient function will depend
on two auxiliary variables $\kappa_i$, whose range (according to the
$\alpha$-representation of the contributing Feynman diagrams) is restricted by
$0\leq \kappa_i \leq 1$.  Here we introduce these variables quite
 trivially through integration over two $\delta$-functions.
In this intuitive way we get the following light-cone expansion:
\begin{eqnarray}
\label{lc1}
\{Tj^\mu(x) j^\nu(0)\}^{\rm as}  &=&  \epsilon^{\mu\nu\rho\sigma}
  \int^1_0 d{\kappa_1}\int^1_0 d{\kappa_2}
   \delta(\kappa_1 -1) \delta(\kappa_2 )  i\partial_\rho^{x} D^c(x)
 \left\{
  O_\sigma(\kappa_1,\kappa_2)+(\kappa_1\leftrightarrow \kappa_2)
 \right\},
\\
\label{p1.6}
  O_\rho(\kappa_1,\kappa_2) &=&
    {\bar\psi}(\kappa_1 \tilde x)\gamma_\sigma\gamma_5 U(\kappa_1
    \tilde x,\kappa_2 \tilde x)   \psi(\kappa_2 \tilde x).
\end{eqnarray}

The light-ray operator $O_\rho$ contains both twist-2 and twist-3
contributions.
In the following we use the light-cone gauge $\tilde{x}A =0 $
so that $U(\kappa_1 \tilde x, \kappa_2 \tilde x) \equiv 1 $.
The twist-2 part of $O_\rho(\kappa_1,\kappa_2)$ can be obtained by
contraction with the light-cone vector $\tilde{x}$:
\begin{eqnarray}
\label{p1.8}
O^{\rm tw 2}(\kappa_1,\kappa_2) = \tilde{x}^\rho O_\rho(\kappa_1,\kappa_2)=
{\bar\psi}(\kappa_1 \tilde x)\not\!\tilde{x} \gamma_5 \psi(\kappa_2 \tilde x).
\end{eqnarray}
It  turns  out  (for  more details see \cite{GeyMueRob96}) that for an
application to forward scattering the twist-3 part can be written as
\begin{eqnarray}
\label{p1.9}
O^{\rm tw 3}_\rho(\kappa_1,\kappa_2) &=&
-i(\kappa_2-\kappa_1)\int_0^1 du\, u\tilde{O}_\rho(\kappa_1,\kappa_1 \bar{u}+
\kappa_2 u),
\nonumber\\
\label{p1.9b}
\tilde{O}_\rho(\kappa_1,\kappa_2) &=&
     \int_0^{1} du\, \bar{\psi}(\kappa_1\tilde{x}) i
     \left[
     \gamma_\rho \tilde{x}_{\sigma}- \not\! \tilde{x}\, g_{\rho\sigma}
\right] \gamma^5
     D^{\sigma}(u,\kappa_1\tilde{x},\kappa_2\tilde{x})
      \psi(\kappa_2 \tilde{x}),
\end{eqnarray}
where we introduced the notation
$D^\rho(u,\kappa_1\tilde{x},\kappa_2\tilde{x})=
    \partial^\rho_{\kappa_2 \tilde{x}}+
  ig A^\rho([\kappa_1 \bar{u}+\kappa_2 u]\tilde{x})$
and $\bar u = 1 - u $.

Using   the   results   of   the   OPE  (\ref{lc1}-\ref{p1.9})  it  is
straightforward  to  obtain  the structure function in terms of parton
distribution  functions.  As  is well known the longitudinal structure
function $g_1$ reads
\begin{eqnarray}
\label{LOg1}
g_1(x_{Bj},Q^2) &=&
 {1\over 2} \sum_{q=u,d,..} e_q^2 \Delta q_q(x_{Bj},Q^2).
\end{eqnarray}
The  quark  distribution  functions  (containing  quark  and antiquark
contributions)  are  defined  as  matrix  elements  of leading twist-2
operators,
\begin{eqnarray}
\label{deftw2DF}
\Delta q_q(x,Q^2)= \int {d\kappa \over 2\pi (\tilde{x}S)}
     \left\langle P,S\left|
       O^{\rm tw 2}(0,\kappa) +
       (\kappa \rightarrow -\kappa)
      \right|P,S \right\rangle
    e^{i x\kappa (\tilde{x}P)},
\end{eqnarray}
where  the  renormalization  point of the operator is set equal to the
momentum transfer $Q^2$. The leading order analysis \cite{WW},\cite{MT}
suggests that the structure function $g_2$ can be written as
\begin{eqnarray}
\label{defg2}
g_2(x,Q^2)=-\bar{g}_2(x,Q^2) + \int_x^1 {dy \over y}\, \bar{g}_2(y,Q^2)
\end{eqnarray}
where   $\bar{g}_2(x,Q^2)   =   g_1(x,Q^2)  +  \tilde{g}_2(x,Q^2)$  is
decomposed into the twist-2 part given by $g_1$ and a remaining twist-3
part $\tilde{g}_2$.
The twist-3 contribution
\begin{eqnarray}
\label{LOg2}
\tilde{g}_2(x_{Bj},Q^2) &=&
 {1\over 2} \sum_{q=u,d,..} e_q^2 \Delta \tilde{q}_q(x_{Bj},Q^2),
\end{eqnarray}
can be expressed by the distribution function
\begin{eqnarray}
\label{deftw3DF}
\Delta \tilde{q}_q(x,Q^2)=
    {1\over x}\int {d\kappa \over 2\pi (\tilde{x}P)} S^\rho
     \left\langle P,S\left|
      \tilde{O}_\rho(0,\kappa) -
      (\kappa \rightarrow -\kappa)
      \right|P,S \right\rangle
    e^{i x\kappa (\tilde{x}P)}.
\end{eqnarray}
where, as  will be seen below, unlike $q_q(x,Q^2)$ this function has no
simple parton interpretation.

\section{Evolution kernels for twist-3 light-ray operators}

The   twist-3   operator   $   \tilde{O}_\rho$  is  not  closed  under
renormalization.  Indeed  it  will  mix  with three-particle operators
which  contain  also  the  gluon field strength. Using the equation of
motion, $(i\!\not\!\!D -m)\psi=0$ the operator $\tilde{O}_\rho$ can be
decomposed with respect to gauge invariant twist-3 operators:
\begin{eqnarray}
\label{p1.15}
\tilde{O}_\rho(\kappa_1,\kappa_2)
    = {i \over 2}(\kappa_2 - \kappa_1)  \int_0^1 du\,
 \big\{\!\!
     &-&\!\! 2 M_\rho(\kappa_1, \kappa_1\bar u +\kappa_2 u) +
  u\; {^+\!S_\rho}(\kappa_1,\kappa_1\bar u+\kappa_2 u,\kappa_2)
\nonumber\\
  &+&\!\!
     \bar u\; {^-\!S_\rho}(\kappa_1,\kappa_1\bar u+\kappa_2u,\kappa_2)
 \big\}+\cdots.
\end{eqnarray}
Here ${^\pm\! S_\rho}$ are the nonlocal generalizations of the so-called
Shuryak-Vainshtein operators  \cite{SHV}:
\begin{eqnarray}
\label{p1.13}
{^\pm\! S_\rho}(\kappa_1,\tau,\kappa_2)&=& ig
    \bar{\psi}(\kappa_1\tilde{x})\not\!\tilde{x}
     \left[
      i \tilde{F}_{\alpha\rho}(\tau\tilde{x}) \pm
      \gamma^5 F_{\alpha\rho}(\tau\tilde{x})
     \right]\tilde{x}^\alpha
    \psi(\kappa_2\tilde{x}),
\end{eqnarray}
where
$\tilde{F}_{\alpha\beta} = {1\over 2} \epsilon_{\alpha\beta\mu\nu} F^{\mu\nu}$
is the dual field strength tensor.
Furthermore,
\begin{eqnarray}
\label{moper}
M_\rho(\kappa_1,\kappa_2)=
        m\; \bar{\psi}(\kappa_1\tilde{x})
     \sigma_{\alpha\rho}\tilde{x}^\alpha \gamma^5
     (\tilde{x}D)(\kappa_2\tilde{x}) \psi(\kappa_2\tilde{x}),
\quad \sigma_{\alpha\beta}={i\over 2} [\gamma_{\alpha},\gamma_{\beta}]
\end{eqnarray}
denotes a mass dependent operator.
Besides the equation of motion operators we neglected also trace terms
(proportional to $\tilde{x}^\rho$) and operators which vanish in the
forward  case (from general principles \cite{JOG} it is known that the
equation  of  motion  operators  do not contribute to the evolution of
physical matrix elements of gauge invariant operators).

For  simplicity  we will neglect the mass operator. Then the operators
${^\pm\!S_\rho}$ which are related to each other by charge conjugation
are  closed  and, furthermore  do  not  mix  under renormalization. We
calculated  the  evolution  kernels  for these operators in light-cone
gauge  using  the  Leibbrandt-Mandelstam prescription \cite{LM}. Since
both  kernels  are  related  to each other we give only the result for
${^-\!S_\rho}$:
\begin{eqnarray}
\label{fresult2}
\mu^2{d\over d\mu^2} {^-\!S^\rho}({\kappa_1},{\kappa_2}) &=&
       {\alpha_s\left(\mu^2\right)\over 4\pi} \int_0^1 dy \int_0^{\bar{y}} dz
 \Big\{ (2C_F-C_A)
    \Big[
     y\,\delta (z)
     {^-\!S^\rho}({\kappa_1}-{\kappa_2}y,-{\kappa_2}y)
\nonumber\\
&&\hspace{0cm}
 -  2z {^-\!S^\rho}(-\kappa_1 y,\kappa_2 -\kappa_1\bar{z})
  +   K(y,z) {^-\!S^\rho}(\kappa_1 \bar{y}  + {\kappa_2}y,
          {\kappa_2}\bar{z}  + {\kappa_1}z) \Big]
\nonumber\\
&&\hspace{-1cm}
+  C_A\Big[
 \Big(2 \bar{z}+ L(y,z) \Big)
     {^-\!S^\rho}({\kappa_1}y,{\kappa_2}-{\kappa_1}z) +
     L(y,z) {^-\!S^\rho}({\kappa_1} - {\kappa_2}z,{\kappa_2}y) \Big]
\Big\},
\end{eqnarray}
\begin{eqnarray}
K(y,z) =
 \left[
  1 + \delta (z)\,{\bar{y}\over y}  + \delta (y)\,{\bar{z}\over z}
 \right]_+,
\quad
L(y,z) =
 \left[
  \delta (1 - y - z)\,{y^2\over \bar{y}} +
 \delta (z)\,{y\over \bar{y}} \right]_+
 - {7\over 4} \delta(\bar{y})\delta(z).
\nonumber
\end{eqnarray}
To condense the notation we used
${^\pm\!S^\rho}(\kappa_1,\kappa_2)={^\pm\!S^\rho}(\kappa_1,0,\kappa_2)$
and the standard plus-prescription
fulfilling $\int dy [...]_+ = 0$ and $\int dy dz [...]_+ = 0$, respectively;
$C_F= (N_c^2-1)/(2N_c)=4/3$ and $C_A=N_c=3$ are the usual Casimir operators
of ${\rm SU}_c(3)$.

\section{Definition of twist-3 parton distribution functions}

Since the twist-3 operators are in fact three-particle operators it is
also  necessary  to  extend  the definition of the  distribution
function  (\ref{deftw3DF}). A quite natural way is to take the Fourier
transform   of   the   operators   ${^\pm\!S^\rho}(\kappa_1,\kappa_2)$
sandwiched  between polarized proton states. The resulting
C-even distribution function
\begin{eqnarray}
\label{defusdisfunc}
\Delta\tilde{Q}^{\rm NS} (x_1,x_2)&=&
  \int {d\kappa_1\over 2\pi} {d\kappa_2\over 2\pi}
   {S_\rho \over 2(\tilde{x}P)^2}
   \langle P,S|
    {^+\!S^\rho}(\kappa_1,\kappa_2) +
    {^-\!S^\rho}(\kappa_2,\kappa_1) +
    \left\{\kappa_1 \leftrightarrow \kappa_2 \right\}
   |P,S\rangle
\\ & &\hspace{5cm}
 \times e^{i\kappa_1 x_1 (\tilde{x}P) + i\kappa_2 x_2 (\tilde{x}P)} 
\nonumber
\end{eqnarray}
depends on the momentum fractions $x_1,x_2$ and posseses the support
property $|x_i|\le 1, |x_1\pm x_2|\le 1$.
Unfortunately, the resulting evolution equation will be very complicated
\cite{BKL,RAT}.
To get a simpler one and to be able to diagonalize it
with respect
to one variable we take into account that operators with different
spin do not mix with each other.

C-even operators with definite spin can be obtained easily from
Shuryak-Vainshtein operators by
\begin{eqnarray}
\label{defYopnonloc}
{Y^{\rho}}(\kappa,u) &=&
    {^+\!S^\rho}(0,\kappa u,\kappa) +
    {^-\!S^\rho}(\kappa,\kappa u,0)
\\
\label{defYoploc}
  Y_n^{\rho}(u) &=&
 i^{n-2} {d^{n-2}\over d\kappa^{n-2}} Y^{\rho}(\kappa,u)_{|\kappa=0}
    =
 \sum_{k=1}^{n-1} {n-2 \choose k-1} u^{k-1} \bar{u}^{n-k-1}
       Y^{\rho}_{n,k},
\end{eqnarray}
The  operators  $Y_n^{\rho}(u)$ possess the definite spin $n$. In Eq.\
(\ref{defYopnonloc})  the  variable  $u$  gives  (for  $\kappa=1$) the
position  of the gluon field on the light cone. For $0\le u \le 1$ the
gluon  field  lies  between  the two quark fields. Note that the gluon
field  can also be outside of this range so that $u$ is not restricted
to the interval $[0,1]$.

Consequently   we  define  a  new  distribution  function  as  Fourier
transform with respect to the variable $\kappa$ only, so that
\begin{eqnarray}
\label{defpart1}
\Delta\tilde{q}^{\rm NS} (y,u) &=&
   \int {d\kappa\over 4\pi}  {S_\rho \over (\tilde{x}P)^2}
   \langle P,S|
    {Y^{\rho}}(\kappa,u)+{Y^{\rho}}(-\kappa,u)
   |P,S\rangle
   e^{i\kappa y (\tilde{x}P)}
\end{eqnarray}
depends on the Fourier conjugate variable $y$ and the gluon position $u$.
To   clarify   the  meaning  of  the  variable  $y$  we  express  this
distribution function (\ref{defpart1})
in terms of the distribution function
$\Delta\tilde{Q}^{\rm NS} (x_1,x_2)$,
\begin{equation}
\label{transform1}
\Delta\tilde{q}^{\rm NS} (y,u)=
 \int_{-1}^1 dx_1\int_{-1}^1 dx_2\;
  \theta(1-|x_1\pm x_2|) \delta(y+u x_1 -\bar{u}x_2)
             \Delta\tilde{Q}^{\rm NS} (x_1,x_2)
\end{equation}
which  gives  the  support  restriction $|y|\leq {\rm Max}(1,|2u-1|)$,
i.e.,  for  $0\le  u\le  1$  the variable $y$ can be interpreted as an
effective momentum fraction.

The  evolution equation for the distribution function (\ref{defpart1})
can    be    derived   from   the   renormalization   group   equation
(\ref{fresult2}).  Using  the transformation
(\ref{defYopnonloc})  and  the definition
(\ref{defpart1})  a straightforward calculation provides the following
result:
\begin{eqnarray}
\label{EvoEqux&u}
Q^2{d\over d Q^2} \Delta\tilde{q}^{\rm NS}\left(y,u,Q^2\right) &=&
 {\alpha_s\left(Q^2\right)\over 2\pi}
 \int {dz\over z} \int dv\,
 \tilde{P}(z,u,v)\,\Delta\tilde{q}^{\rm NS}\left({y\over z},v,Q^2\right).
\end{eqnarray}
Here the integration region is determined by both the support of
$\Delta\tilde{q}^{\rm NS}$ and the support of the evolution kernel
\newpage
\begin{eqnarray}
\label{EvoKer}
&&\tilde{P}_{qq}(z,u,v) =
\\
&&{2C_F-C_A \over 2}
\Bigg[
  [\Theta_1(z,u,v) K(z,u,v)]_+  + 
 \Theta_2(z,\bar{u},\bar{v}) L(z,\bar{u},\bar{v})-\Theta_2(z,u,v) M(z,u,v)
\Bigg]+
\nonumber\\
&&\hspace{0cm}
 {C_A \over 2}\left[
 \left\{
   [\Theta_3(z,u,v) N(z,u,v)]_+ +
     {u\rightarrow \bar{u} \choose v\rightarrow \bar{v}}
 \right\}+
 \Theta_3(z,u,v) M(z,u,v)-{7\over 2} \delta(u-v)\delta(1-z)\right],
\nonumber
\end{eqnarray}
where
$[A(z,u,v)]_+=A(z,u,v)-\delta(1-z)\delta(u-v)
    \int_0^1 dz'\int dv'\, A(z',u,v')$
and
the auxiliary functions are defined by:
\begin{eqnarray}
\label{auxfunc}
\Theta_1(z,u,v)&=&
  \theta(z) \theta(u-z v) \theta(\bar{u}-z\bar{v}),
\qquad
\Theta_2(z,u,v)=
 \theta(-\bar{u}\bar{v}z) \theta(\{1-v z\}\bar{u})
 \theta(\{z-u\}\bar{u}),
\nonumber\\
\Theta_3(z,u,v)&=&
 \theta(\bar{u}\bar{z})\theta(\bar{u}\bar{v}z)\theta(\{v z-u\}\bar{u}),
\nonumber\\
K(z,u,v) &=&
  z+
 \left\{
   {u^2 \over v (v-u)}\delta(u-z v) +
  {u\rightarrow \bar{u} \choose v\rightarrow \bar{v}}
 \right\},
\quad
L(z,u,v) =
 -{\rm sign}(\bar{u}){\bar{v} u^2 \over \bar{u}^2} \delta(u-z),
\nonumber\\
M(z,u,v) &=&
  {2z(1-z v) \over \bar{u}^3},
\quad
N(z,u,v) =
  {{\rm sign}(u) \bar{v}\over \bar{u} (v-u)}
   \left\{
    {\bar{v}\over \bar{u}} \delta(1-z) +
    {u^2\over v} \delta\left(u-z v\right)
   \right\}.
\end{eqnarray}

\section{Solution of the evolution equation}

A diagonalization of the evolution equation (\ref{EvoEqux&u})
can be achieved partly by the Mellin transform
\begin{eqnarray}
 \Delta\tilde{q}_n^{\rm NS} (u) =
 \int_0^1 dx\, x^{n-2}  \Delta\tilde{q}^{\rm NS} (x,u).
\end{eqnarray}
These moments are given by matrix elements of the operators $Y_n^{\rho}(u)$
with spin $n$ and are polynomials in the variable $u$ of order $(n-2)$
[see Eq.\ (\ref{defYoploc})].
Their evolution equation reads
\begin{eqnarray}
\label{EvolMom}
Q^2{d\over dQ^2} \Delta\tilde{q}_n^{\rm NS} \left(u,Q^2\right) &=&
 {\alpha_s\left(Q^2\right)\over 2\pi}
 \int dv\,
 \tilde{P}_n(u,v)\, \Delta\tilde{q}_n^{\rm NS}\left(v,Q^2\right),
\end{eqnarray}
where the  $n$-dependent kernel
$\tilde{P}_n(u,v)= \int_0^1 dz\, z^{n-2} \tilde{P}_n(z,u,v)$ can be
calculated from Eq.\ (\ref{EvoKer}). For each given $n$,
$\Delta  \tilde{q}_n^{\rm NS} (u)$ can be expanded with respect to the
$(n-2)$  eigenfunctions $e_{n,i}(u)$ of the kernel $\tilde{P}_n(u,v)$.
These polynomials can be constructed by diagonalization of
\begin{eqnarray}
\label{reltolocanomdim}
 {n-2\choose j-1} \int dv\,\tilde{P}^n_{qq}\left(u,v\right)
         v^{n-j-1} \bar{v}^{j-1} &=&
 \sum_{i=1}^{n-1} {n-2\choose i-1}
   \tilde{\gamma}_{ij}^n\, u^{n-i-1} \bar{u}^{i-1},
\end{eqnarray}
where  $\tilde{\gamma}_{ij}^n$  are  the known anomalous dimensions of
the   local   operators   ${Y}^\rho_{n,i}$  \cite{BKL,KO}.  A  general
analytical  solution  of this problem is not known. But, for the first
few moments the numerical solution can be obtained easily.

Knowing  the  $(n-2)$  eigenfunctions  $e_{n,i}(u)$  and eigenvalues $
\lambda_{n,i}$   of  the  kernel  $\tilde{P}_n(u,v)$  it  is  possible
to determine  the evolution of the moments
\begin{eqnarray}
\label{solution}
\Delta\tilde{q}_n^{\rm NS} \left(u,Q^2\right) =
 \sum_{i=1}^{n-1} c_i\left(Q_0^2\right) e_{n,i}(u)
 \exp{
  \left\{
  \int_{Q_0^2}^{Q^2} {dt\over t} {\alpha_s(t)\over 2\pi} \lambda_{n,i}
  \right\}}.
\end{eqnarray}
The   nonperturbative   coefficients   $c_i\left(Q_0^2\right)$   at  a
reference  momentum  squared  $Q^2_0$  have  to be determined from the
initial   value   $\Delta\tilde{q}_n^{\rm  NS}  \left(u,Q_0^2\right)$.
Since,  assuming  the  validity  of  Eq.\  (\ref{defg2}),  the twist-3
nonsinglet part of the structure function $g_2$ satisfies
\begin{eqnarray}
\label{reltog2}
x \tilde{g}_2^{\rm NS}(x) = {d\over dx} \Delta\hat{q}^{\rm NS} (x),\quad
\Delta\hat{q}^{\rm NS}(x,Q^2) =
  \int_0^1 du\, u \Delta\tilde{q}^{\rm NS}(x,u,Q^2)
\end{eqnarray}
it is obvious that the initial values can not be obtained alone from
transversal polarized deep inelastic scattering experiments.

Finally,  we show that the above mentioned complications do not appear
in  the  limit  $x\to  1$  as well as in limit of large $N_C$. We deal
directly with $\Delta\hat{q}^{\rm NS} (x)$. From the definition of the
evolution  kernel  (\ref{EvoKer})  it  turns out that $\int_0^1 du\, u
\tilde{P}(z,u,v)$  factorizes  in  boths limits according to $v P(z)$.
Thus,  from the evolution equation (\ref{EvoEqux&u}) follows for $x\to
1$,
\begin{eqnarray}
\label{soleveqxto1}
&& Q^2{d\over dQ^2} \Delta\hat{q}^{\rm NS}\left(x,Q^2\right) =
\\
 &&\hspace{1 cm}
    {\alpha_s\left(Q^2\right)\over 2\pi} \int_x^1 {dz\over z}
 \left\{\left[2C_F\over 1-z \right]_+
 +\left({3 \over 2} C_F-C_A\right)\delta(1-z)\right\}
    \Delta\hat{q}^{\rm NS}\left({x\over z},Q^2\right) + O\left({1-x}\right),
\nonumber
\end{eqnarray}
and for large $N_c$,
  \begin{eqnarray}
\label{soleveqNc}
&& Q^2{d\over dQ^2} \Delta\hat{q}^{\rm NS}\left(x,Q^2\right) =
\\&&\hspace{1 cm}
 {\alpha_s\left(Q^2\right)\over 4\pi}N_c  \int_0^1 {dz\over z}
  \left\{
   2\left[z^2\over 1-z \right]_+ -z^2 -{5\over 2} \delta(1-z)
         \right\}
 \Delta\hat{q}^{\rm NS}\left({x\over z},Q^2\right) +
O\left({1\over N_c}\right).
\nonumber
\end{eqnarray}
These equations are  of the GLAP type and coincide with the result of
\cite{AliBraHil91}. The $1/N_c$ suppressed terms contain contributions from
the region $v<0$ and $v>1$ and they provide a system of coupled evolution
equations.

\section*{Acknowledgment}

We   wish   to   thank   J.~Bl\"umlein,   V.~M.~Braun,   E.~A.~Kuraev,
L.~N.~Lipatov,  and  O.~V.~Teryaev for valuable discussions. D.~M. was
financially supported by Deutsche Forschungsgemeinschaft (DFG) and the
Landau-Heisenberg program.


\newpage

\end{document}